\documentclass[10pt]{article}

\usepackage{amsmath}
\usepackage{array}
\usepackage{appendix}
\usepackage{graphicx}
\usepackage{amsfonts}
\usepackage{amssymb}
\usepackage{mathrsfs}
\usepackage{yfonts}
\usepackage{euscript}
\usepackage{upgreek}
\usepackage{slantsc}
\usepackage{calligra}
\usepackage[T1]{fontenc}
\usepackage{epsf}
\usepackage{latexsym}

\usepackage{tipa}

\def\be{\begin{equation}}
\def\ee{\end{equation}}
\def\beq{\begin{equation}}
\def\eeq{\end{equation}}
\def\bea{\begin{eqnarray}}
\def\eea{\end{eqnarray}}

\def\ni{\noindent}

\def\hat{\widehat}

\def\!{\hspace{-1.6667em}}

\def\mD{\mbox{D}}

\def\mathfrakV{\mbox{\LARGE $\mathfrak{v}$}}

\def\FrQ{\mbox{\Large $\mathfrak{q}$}}
\def\FrK{\mbox{\boldmath$\mathfrak{K}$}}                   
\def\FrN{\mathfrak{N}}                            

\def\FrH{\mbox{\boldmath$\mathfrak{H}$}}          

\def\sFrG{\mbox{\small$\mathfrak{g}$}}

\def\FrG{\mbox{\Large $\mathfrak{g}$}}  
 
\def\sa{\mbox{\scriptsize a}}

\def\scc{\mbox{\scriptsize c}}

\def\se{\mbox{\scriptsize e}}

\def\sn{\mbox{\scriptsize n}} 
\def\so{\mbox{\scriptsize o}}

\def\sv{\mbox{\scriptsize v}}

\def\tn{\mbox{\tiny n}}

\def\pa{\partial}
\def\d{\textrm{d}}

\def\5Star{\mbox{\Large$\star$}}              
\def\sumi2{\sum\mbox{}_{\mbox{}_{\mbox{\scriptsize $i$=1}}}^2}
\def\sumi3{\sum\mbox{}_{\mbox{}_{\mbox{\scriptsize $i$=1}}}^3}

\def\sumj3{\sum\mbox{}_{\mbox{}_{\mbox{\scriptsize $j$=1}}}^3}
\def\sumk3{\sum\mbox{}_{\mbox{}_{\mbox{\scriptsize $k$=1}}}^3}

\begin{document}

\begin{center}

{\bf \Large KINEMATICAL QUANTIZATION}

\vspace{.15in}

{\large \bf Edward Anderson} 

\vspace{.15in}

\large {\em DAMTP, Centre for Mathematical Sciences, Wilberforce Road, Cambridge CB3 OWA.  } \normalsize

\end{center}

\begin{abstract}

We consider here kinematical quantization: a first and often overlooked step in quantization procedures.
$\mathbb{R}$, $\mathbb{R}_+$ and the interval are considered, as well as direct (Cartesian) products thereof.
Some simple minisuperspace models, and mode by mode consideration of slightly inhomogeneous cosmology, have indefinite signature versions of such kinematical quantizations.
The examples in the current paper build in particular toward the case of vacuum $\mathbb{S}^3$ slightly inhomogeneous cosmology's mode configuration space, 
which is mathematically a finite time interval slab of Minkowski spacetime.  

\end{abstract}

\section{Introduction}

Consider a classical system with configuration variables $Q^A$, the range of possible values for which form the configuration space $\FrQ$ \cite{AConfig}.  
The corresponding conjugate momenta are denoted by $P_A$.
The $Q^A$, the $P_A$ and the Poisson bracket $\mbox{\bf \{} \mbox{ } \mbox{\bf ,} \mbox{ } \mbox{\bf \}}$ constitute the phase space of the system.\footnote{`Symplectic form' \cite{Arnold}
can also be evoked at this stage in place of Poisson bracket.
On some occasions, the Dirac bracket comes to be in use instead of the Poisson bracket due elimination of second-class constraints \cite{Dirac, HT92}.
However, this can also be regarded as a reduced geometrical system's own notion of Poisson bracket \cite{Sni}.
See also \cite{Souriau, Arnold, HT92, BojoBook} for more on symplectic and Poisson manifolds.}
%
Classical functions $f(Q^A, P_A)$ can be considered without problems: all such can be considered together without any restrictions.
It is more demanding to ask for such which classical brackets commute with a theory's first-class constraints.
I.e. the classical `Problem of Observables' \cite{Dirac, HT92, PoT1, PoB}, which is particularly problematic in the case of Gravitational Theory.

\mbox{ }

\ni Quantum-level functions, however, are subject to restrictions.
This is closely related to kinematical quantization \cite{I84}, 
a step which few physicists even notice due to the answer to its issues in common cases being ubiquitously known and taken as a given.  
E.g. that one has $\hat{\underline{Q}}$, $\hat{\underline{P}}$ and $\hat{\underline{J}}$ for a particle in 3-$d$. 
But what is the analogue of these for a more general quantum system's quantization? 

\mbox{ }

\ni The classical Poisson brackets subalgebraic structure leads to some commutator algebra, 
$\widehat{f}(\widehat{Q^{A}}, \widehat{P}_{A}) \mbox{ that close under } \mbox{\bf [} \mbox{ }  \mbox{\bf ,} \mbox{ }  \mbox{\bf ]}$, which is far from necessarily isomorphic to it.
The latter comes with commutator-preserving morphisms $M$ in place of the classical-level canonical transformations which preserve the Poisson brackets.
The space of operators in question in this case follows from the kinematical operators having a space of wavefunctions to act upon.
This is mathematically a Hilbert space, Hilb.  
However, it is not the particular physically realized Hilbert space, due to dynamical (and in the next Chapter $\FrG$) considerations having not yet been made.
Due to this, I denote Hilbert spaces in the former role by KinHilb for `{\it kinematical Hilbert space}', and those in the latter role by DynHilb for `{\it dynamical Hilbert space}'; 
the current article, however, only considers KinHilb in further detail.   
The corresponding morphisms are unitary transformations, $Uni$.  

\mbox{ }

\ni Next, returning to the restrictions arising at the quantum level, one source of such is the {\it Groenewold--van Hove phenomenon} \cite{Gotay00}.  
By this a wide range of sets of functions of variables cannot be allotted commutator brackets in a consistent manner.
This leaves one having to select a preferred set of classical $f(Q^{A}, P_{A})$ that are to be promoted to quantum operators $\widehat{f}(\widehat{Q}^{A}, \widehat{P}_{A})$. 
The preferred set, a fortiori, need to algebraically close under the classical Poisson bracket $\mbox{\bf \{} \mbox{ } \mbox{\bf ,} \mbox{ } \mbox{\bf \}}$, 
and so form a subalgebraic structure. 
The operators these are promoted to then also need to close under the quantum commutator bracket (this is not implied by the previous sentence for reasons laid out in \cite{ABrackets}).  

On the other hand, the latter are to algebraically close under the commutator bracket $\mbox{\bf [} \mbox{ } \mbox{\bf ,} \mbox{ } \mbox{\bf ]}$.

\mbox{ } 

\ni Kinematical quantization can be viewed as a first step in canonical quantization, 
in particular one which is well adapted to `deformation' \cite{Weyl27-GVH1} or `geometrical' \cite{Souriau, Kostant-WoodhouseBook, A91} approaches to quantization; 
see e.g. \cite{Landsman} for some further developments in these directions.
The current Article follows a configuration space geometry centred method exposited by Isham \cite{I84}.  
The current article illustrates with examples how QM has greater sensitivity to global structure (Secs 2 and 3).  
General features in this regard are then as follows. 

\mbox{ }

\ni 1) The operators are to respect the underlying configuration space's limitations in extent.

\mbox{ }

\ni 2) The operators are to respect the basic self-adjointness criterion \cite{RS}. 

\mbox{ }

\ni 3) The selected operators are to close as an algebraic structure under the quantum mechanical commutator brackets.   

\mbox{ }

\ni 4) Let us next consider a method which works for quite a wide range of examples.
This is due to Mackey \cite{Mackey} and was widely used in the construction of examples by Isham \cite{I84, IConcat, I-Cat}.  
It involves a configuration space of the form of a homogeneous space $\FrG_1/\FrG_2$ for $\FrG_2$ a subgroup of $\FrG_1$. 

\mbox{ } 

In this case, the corresponding kinematical quantum algebraic structure $\FrK$ can be decomposed as semidirect products\footnote{The semidirect product $\sFrG = \FrN \rtimes \FrH$ 
is given by $(n_1, h_1) \circ (n_2, h_2) = (n_1 \circ \varphi_{h_1}(n_2), h_1\circ h_2)$ for $\langle \FrN, \circ \rangle  \lhd \sFrG$, 
                                                                                   $\langle \FrH, \circ \rangle$ a subgroup of $\sFrG$ 
                                                                                       and $\varphi:\FrH \rightarrow \mbox{Aut}(\FrN)$ a group homomorphism.}
\beq
\mathfrakV^*(\FrQ) \rtimes \FrG_{\scc\sa\sn}(\FrQ) \mbox{ } . 
\eeq
Here, $\FrG_{\scc\sa\sn}(\FrQ)$ is the {\it canonical group} and $\mathfrakV^{*}$ is the dual of a linear space $\mathfrakV$.  
This is natural due to carrying a linear representation of $\FrQ$ with the property that there is a $\FrQ$-orbit in $\mathfrakV$ is diffeomorphic to $\FrQ/\FrG$ \cite{I84}.  
Then 
\be
\FrK = (\mathfrakV^{*} \rtimes \FrG_{\scc\sa\sn}(\FrQ), \mbox{\bf [} \mbox{ } \mbox{\bf ,} \mbox{ }\mbox{\bf ]}) \mbox{ } .
\ee
Also $\mathfrakV^* = \mathfrakV$ for finite examples, while $\FrG_{\scc\sa\sn}(\FrQ) = Isom(\FrQ)$ -- the corresponding isometry group -- for some of the examples in this Article. 
As regards the passage to the corresponding Representation Theory -- which plays a major role at the quantum level -- 
for semidirect products the powerful techniques of Mackey Theory \cite{Mackey} apply.  

\mbox{ } 

\ni 5) There are extra pieces in the case of nontrivial cocycles.\footnote{Cohomology plays a very important role in more advanced considerations of quantization.
See \cite{IshamBook2} for an outline of cohomology, 
\cite{Cohom} for more details, 
\cite{Nash, HT92} for outlines of some physical applications, 
and \cite{I84} for the current Article's specific application.}  

\mbox{ }

\ni While the current Article's account of $\mathbb{R}_+$ is but a useful recollection, the Article's goal is to build on this toward the case of a slab of flat indefinite space 
as occurs in the leading order perturbative modewise treatment of slightly inhomogenous cosmology \cite{HallHaw, SIC-1, SIC-2}. 
This is an interesting quantum cosmological model as regards the origin of galaxies and cosmic microwave background inhomogeneities, 
and also a minimally sufficient model \cite{SIC-1} as regards exhibiting all of the facets of the Problem of Time \cite{PoT1, PoT2}.
The models intermediate between this and $\mathbb{R}$, $\mathbb{R}_+$ are furthermore useful in relational mechanics \cite{BB82B03FORD, FileR, AConfig, AMech} 
and in minisuperspace \cite{Magic, Ryan-BI75-HH83-AMSS1}, which are both also useful model arenas for Quantum Cosmology.

\section{$\mathbb{R}$ and $\mathbb{R}_+$}\label{Sec-2}

For a particle in $\mathbb{R}$, the conventional selection is just $x$, $p$, 
followed by the promotion  $q \longrightarrow \widehat{x}$, $p \longrightarrow \widehat{p}$ that can be represented by $\widehat{x} = x$ and $\widehat{p} = -i\hbar \, \pa/\pa x$.
These are self-adjoint in the obvious
\beq 
\mbox{KinHilb} = L^2
\left(
\mathbb{R}, \d x
\right)                                                                \mbox{ } ,
\eeq
This case's commutation relation's 1 or $i\hbar$ term is, moreover, the result of a nontrivial cocycle, resolved by admitting a central extension \cite{I84}.  
This is why $Heis(1)$ contains three pieces:  $p$, $x$ and $1$.
In this case, the classical Poisson brackets and quantum commutator structures are an isomorphic pair,  
\beq
\mbox{from } \mbox{ } \mbox{\bf\{}x \mbox{\bf ,} \, p \mbox{\bf\}} = 1            \mbox{ } \mbox{ to } \mbox{ } 
\mbox{\bf [}\widehat{x} \mbox{ } \mbox{\bf ,} \, \widehat{p}\mbox{\bf ]} = i\hbar \mbox{ } ;   
\label{common}
\eeq
the nontrivial cocycle causes no change of algebraic structure since in this case it is already present at the classical level.

\mbox{ }

\ni Isham \cite{I84} also considered the case of $\mathbb{R}_+$.  
Suppose one were to try to represent $x$ and $p$ by 
\be
\widehat{x} = x \mbox{ } , \mbox{ } \mbox{ }  \widehat{p} = -i \,\hbar \, \pa/\pa x
\ee
here. 
Then the latter is not essentially self-adjoint since it does not respect the endpoint of the $\mathbb{R}_+$ by continuing to generate a translation past it (Fig \ref{R+mom}.a).  
To avoid this, one uses instead the representation 
\be
\widehat{p} = -i \, \hbar \, x \pa/\pa x
\ee 
(Fig \ref{R+mom}.b).
Moreover, the quantum commutator is then
\beq
\mbox{\bf[}\widehat{x} \mbox{\bf ,}  \, \widehat{p} \mbox{\bf ]} = i \, \hbar \, \widehat{x} 
\eeq
-- the {\it affine} commutation relation.
Thus this example illustrates a distinct algebraic structure applying at the quantum level 
[the corresponding classical Poisson bracket still being of the standard fundamental bracket form].
This change is due to Quantum Theory's greater sensitivity to topological structure; in particular, while $\mathbb{R}_+$ is contractible, it is however not a vector space.  
In this case, 
\be
\mbox{KinHilb} = L^2
\left(
\mathbb{R}_+, \frac{\d x}{x}
\right)                       \mbox{ } ;
\ee 
in particular, the constructed $\widehat{p}$ is self-adjoint with respect to this and not with respect to the `usual' measure $\d x$.  

{            \begin{figure}[ht]
\centering
\includegraphics[width=0.7\textwidth]{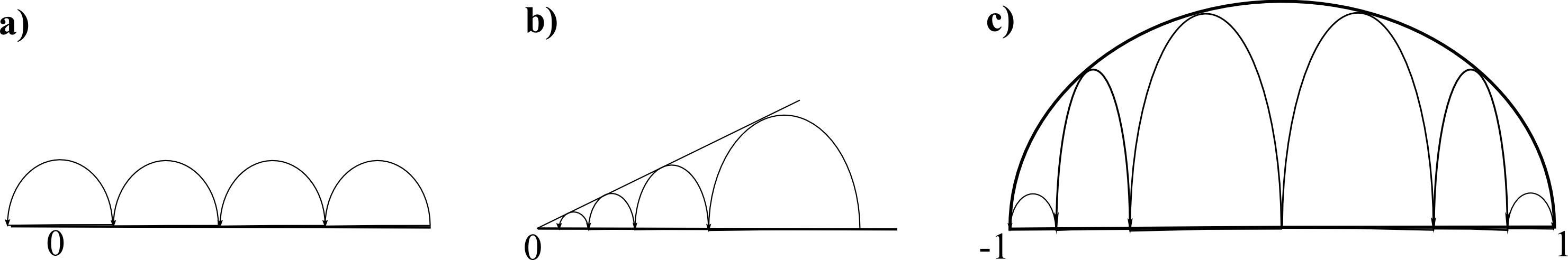}
\caption[Text der im Bilderverzeichnis auftaucht]{        \footnotesize{Heuristic picture of a) unsuitable and b) suitable operator actions as regards representing momentum on the real half-line.
c) Suitable operator actions as regards representing momentum on the interval.
}   }
\label{R+mom} 
\end{figure}  } 

\ni Note that the $\mathbb{R}$ case's fundamental Poisson bracket's constant right hand side term can be interpreted as an obstruction 2-cocycle \cite{I84} 
whose presence necessitates a central extension; this interpretation adds insight to the subsequent quantization procedure.
By this e.g. the $\FrQ = \mathbb{R}$ case's $\FrK$ picks up an extra $\mathbb{R}$ -- the 1 -- 
in addition to the $\mathbb{R}$ of $q$ in the $\mathfrakV$ and the $\mathbb{R}$ of $p$ which forms $\FrG_{\scc\sa\sn}(\mathbb{R})$.  
In contrast, the $\mathbb{R} \rtimes \mathbb{R}_+$ case involves no central extension term. 
Furthermore, this case also does not involve $Isom(\mathbb{R}_+)$ either. 
This is because $\pa/\pa x$ is the Killing vector which forms the 1-$d$ isometry group.  
On the one hand, $\pa/\pa x$ arises from locally solving the Killing equation, 
but on the other hand this expression does not comply with the requisite essential self adjointness to be part of $\FrK$.

\mbox{ } 

\ni Let us now return to the $\mathbb{R}_+$ example. The Laplace-ordered \cite{DeWitt57} time-independent Schr\"{o}dinger equation\footnote{This is often used in Quantum Cosmology, 
though the reasons for this apply to a wider range of operator orderings, among which the conformal ordering may be even better \cite{Magic, PPSCT}.
The two coincide in some cases though, e.g. on flat spaces and in 2-$d$, and the Laplace ordering gives slightly simpler equations, so it is often used in schematic considerations.} 
is then 
\beq
-\frac{\hbar^2}{2} x \frac{\d}{\d x}\left\{ x \frac{\d}{\d x} \right\}\Psi + V(x)\Psi = E\Psi
\eeq
E.g. for the free case this is a Bessel equation, and for the case with a power-law potential it can be mapped to the associated Laguerre equation.
Also note that this equation differs from the $\mathbb{R}$ case's more familiar 
\beq
-\frac{\hbar^2}{2} \frac{\d^2}{\d x^2} \Psi + V(x)\Psi = E\Psi \mbox{ } ,
\eeq
including as regards the former needing two more powers in the potential so as to have boundedness.
This illustrates kinematical quantization imprint on wave equations.
In this way, it is a first simple model arena of the distinction between plain and affine \cite{Affine} 
Wheeler--DeWitt equations (I.e. the at least apparently time-independent wave equations which occur in canonical GR).

\section{The interval}\label{Sec-3}

For an interval, there are two endpoints to respect. 
Without loss of generality, place these at $\pm 1$. 
Then 
\beq
x \mbox{ } , \mbox{ } \mbox{ } \sqrt{1 - x^2} \mbox{ and } -i \, \hbar \, \sqrt{1 - x^2}\pa/\pa x  
\eeq
are self-adjoint in 
\beq 
\mbox{KinHilb} = L^2
\left(
\mathbb{R}_+, \frac{\d x}{\sqrt{1 - x^2}}
\right)                                                                \mbox{ } ,
\eeq
and close as  
\beq
\mbox{\bf [} p \mbox{\bf ,} \, x \mbox{\bf ]}              = - i\hbar \sqrt{1 - x^2} \mbox{ } , \mbox{ } \mbox{ } 
\mbox{\bf [} p \mbox{\bf ,} \, \sqrt{1 - x^2} \mbox{\bf ]} =   i\hbar    \,    x      \mbox{ } . 
\eeq
This can be identified as the Euclidean algebra
\beq
Eucl(2) = \mathbb{R}^2 \rtimes SO(2) \mbox{ } . 
\eeq
Furthermore identifying 
\be
x = \mbox{cos}\, \theta \mbox{ } ,
\ee 
this can be more familiarly re-represented as 
\be
\mbox{sin} \, \theta        \mbox{ } ,  \mbox{ }  \mbox{ } 
\mbox{cos} \, \theta        \mbox{ } \mbox{ and } \mbox{ } 
- i \, \hbar \, \pa/\pa \theta \mbox{ } ,
\ee 
though these receive a different interpretation here than in the case of $\mathbb{S}^1$ \cite{I84}, due to --1 and 1 now not being identified.  
\ni The Laplace-ordered time-independent Schr\"{o}dinger equation is then 
\beq
-\frac{\hbar^2}{2} \sqrt{1 - x^2} \frac{\d}{\d x}\left\{ \sqrt{1 - x^2} \frac{\d}{\d x} \right\}\Psi + V(x)\Psi = E\Psi
\eeq
E.g. in the free case, this is a Tchebychev equation, 
for which the second solution goes like
\be
\mbox{sin}(n \, \mbox{arccos} x) = \sqrt{1 - x^2} \, \mbox{sin}(n x)/\mbox{sin}\, x \mbox{ } , 
\ee 
which succeeds in meeting the boundary conditions of vanishing at $\pm 1$.

\section{Direct (Cartesian) product composites}

\ni For a particle in $\mathbb{R}^2$, the kinematical quantization scheme's view is that the selection does not just involve $x^i$ and $p_i$ but $J$ as well. 
Upon promoting these to quantum operators, they can be represented by 
\be
\widehat{x}^i = x^i, \widehat{p}_i = -i \, \hbar \, \pa/\pa x^i \mbox{ and } \widehat{J} = - i \, \hbar \{y \pa/\pa x - x \pa/\pa y\} = -i \, \hbar \, \pa/\pa \phi \mbox{ } .  
\ee
\ni For a particle in $\mathbb{R}^3$, the kinematical quantization scheme's view is that the selection is not just $x^i$ and $p_i$ but $J_i$ as well. 
Upon promoting these to quantum operators, they can be represented by 
\be
\widehat{x}^i = x^i                                                 \mbox{ } , \mbox{ }  \mbox{ } 
\widehat{p}_i = -i \, \hbar \, \pa/\pa {x^i}                        \mbox{ and }         \mbox{ } 
\widehat{J}_i = -i \, \hbar \, \epsilon_{ijk}x^j \pa/\pa {x^k}      \mbox{ } .
\ee  
\ni The 3-$d$ angular momenta can also be cast in dual form 
\be
J_{ij} = i \, \hbar\{x_i \pa/\pa x_j - x_j \pa/\pa x_i\} \mbox{ } ,
\ee 
which presentation extends to $n$-$d$.
In this case, the classical Poisson brackets and quantum commutator structures are an isomorphic pair for each $d$. 
Moreover, at the classical level one might well consider Poisson brackets of classical quantities other than the $x$'s $p$'s and $J$'s, 
or not think to allot co-primary status to the $J$'s.
We shall see below that kinematical quantization is more precise in this regard, allotting co-primary status to the $\widehat{x}$'s, $\widehat{p}$'s and $\widehat{J}$'s.
Finally, for $\mathbb{R}^n$, 
\be 
\mbox{KinHilb} = L^2 
\left( 
\mathbb{R}^n, \prod_{i = 1}^n \d x^i 
\right) \mbox{ } . 
\ee
The Heisenberg group $Heis(n)$ arising in kinematical quantization involves `the unit' $\delta_{ij}$ and the $x_i$ in addition to the $Eucl(n)$'s $p_i $ and $J_{ij}$.  

\mbox{ }

\ni Example 1) $\mathbb{R}_+^n$ has 
\be
x_i                   \mbox{ } , \mbox{ } \mbox{ } 
x_i \pa/\pa x_j           \mbox{ } ,
\ee
which are self adjoint with respect to 
\be
\mbox{KinHilb} = L^2
\left(
\mathbb{R}_+^n, \frac{\prod_i \d x_i}{x_i} 
\right)                                     \mbox{ } ,
\ee 
and form the $n$-$d$ affine algebra 
\be
Aff(n) = \mathbb{R}^n \rtimes GL^+(n, \mathbb{R}) \mbox{ } .
\ee
\ni Example 2) Kinematical quantization of $\mathbb{R} \times \mathbb{R}_+$ involves 
\be
x                       \mbox{ } , \mbox{ } \mbox{ }  
y                       \mbox{ } , \mbox{ } \mbox{ }  
     \pa/\pa x          \mbox{ } , \mbox{ } \mbox{ }  
y \, \pa/\pa y          \mbox{ and }        \mbox{ } 
y \, \pa/\pa x          \mbox{ } .
\ee  
These are all self-adjoint with respect to 
\be
\mbox{KinHilb} = L^2
\left(
\mathbb{R} \times \mathbb{R}_+, \frac{\d x \, \d y}{y}
\right)
\ee 
and respect approach to $y = 0$ end of range of $\FrQ$.  
This 5-$d$ algebra closes as 
\be
\mbox{\bf [}      \pa/\pa x \mbox{\bf ,} \, x              \mbox{\bf ]} = 1               \mbox{ } , \mbox{ } \mbox{ } 
\mbox{\bf [} y \, \pa/\pa y \mbox{\bf ,} \, y              \mbox{\bf ]} = y               \mbox{ } , \mbox{ } \mbox{ } 
\mbox{\bf [} y \, \pa/\pa x \mbox{\bf ,} \, x              \mbox{\bf ]} = y               \mbox{ } , \mbox{ } \mbox{ } 
\mbox{\bf [} y \, \pa/\pa y \mbox{\bf ,} \, y \, \pa/\pa x \mbox{\bf ]} = y \, \pa/\pa x \mbox{ } .  
\ee
\ni Example 3) Extending the above to $\mathbb{R}^n \times \mathbb{R}_+$ involves 
\be
x_i                                      \mbox{ } , \mbox{ } \mbox{ } 
y                                        \mbox{ } , \mbox{ } \mbox{ } 
\pa/\pa {x_i}                            \mbox{ } , \mbox{ } \mbox{ } 
y \, \pa/\pa y                           \mbox{ } , \mbox{ } \mbox{ } 
x_j  \pa/\pa {x_i} - x_i \pa/\pa {x_j}   \mbox{ and }        \mbox{ } 
y \, \pa/\pa {x_i}                       \mbox{ } . 
\ee
These are all self-adjoint with respect to 
\be
\mbox{KinHilb} = L^2
\left(
\mathbb{R}^m \times \mathbb{R}, \frac{\prod_i\d x_i \d y}{y}  
\right)
\ee 
and respect approach to $y = 0$ end of range of $\FrQ$.  
This $\Big(n(n + 5)/2 + 2\Big)$-$d$ algebra closes as 
\be
\mbox{\bf [}      \pa/\pa x_i \mbox{\bf ,} \, x_j              \mbox{\bf ]} = \delta_{ij}   \mbox{ } , \mbox{ } \mbox{ } 
\mbox{\bf [} y \, \pa/\pa y   \mbox{\bf ,} \, y                \mbox{\bf ]} = y             \mbox{ } , \mbox{ } \mbox{ } 
\mbox{\bf [} y \, \pa/\pa x_i \mbox{\bf ,} \, x_j              \mbox{\bf ]} = y \delta_{ij} \mbox{ } , \mbox{ } \mbox{ } 
\mbox{\bf [} y \, \pa/\pa y   \mbox{\bf ,} \, y \, \pa/\pa x_i \mbox{\bf ]} = y \pa/\pa x_i   \mbox{ } .  
\ee
\ni Example 4) Kinematical quantization of $\mathbb{R} \times I$ involves the algebra $\FrK$ of seven elements
\beq
x                                  \mbox{ } , \mbox{ } \mbox{ } 
y                                  \mbox{ } , \mbox{ } \mbox{ } 
\sqrt{1 - y^2}                     \mbox{ } , \mbox{ } \mbox{ } 
\pa/\pa x                          \mbox{ } , \mbox{ } \mbox{ } 
\sqrt{1 - y^2} \pa/\pa y           \mbox{ } , \mbox{ } \mbox{ }  
\sqrt{1 - y^2} \pa/\pa x           \mbox{ }   \mbox{  and } \mbox{ } \mbox{ }  
y \, \pa/\pa x                     \mbox{ } .
\eeq
\ni These are all self-adjoint with respect to 
\be
\mbox{KinHilb} = L^2
\left(
\mathbb{R} \times I, \frac{\d x \, \d y}{\sqrt{1 - y^2}}
\right)
\ee 
and respect approach to the $y = \pm 1$ end of range of $\FrQ$.  

\mbox{ }

\ni Example 5) Extending to $\mathbb{R}^n \times I$, kinematical quantization involves the algebra $\FrK$ of $n(n + 7)/2 + 3$ elements 
\be
x_i                                     \mbox{ } , \mbox{ } \mbox{ }  
y                                       \mbox{ } , \mbox{ } \mbox{ }  
\sqrt{1 - y^2}                          \mbox{ } , \mbox{ } \mbox{ }  
\pa/\pa {x_i}                           \mbox{ } , \mbox{ } \mbox{ }  
\sqrt{1 - y^2}\pa/\pa y                 \mbox{ } , \mbox{ } \mbox{ } 
\sqrt{1 - y^2}\pa/\pa {x_i}             \mbox{ } , \mbox{ } \mbox{ }  
y \, \pa/\pa {x_i}                      \mbox{ and }        \mbox{ } 
x_j\pa/\pa {x_i} - x_i\pa/\pa {x_j}     \mbox{ } .
\ee
These are all self-adjoint with respect to 
\be
\mbox{KinHilb} = L^2
\left(
\mathbb{R}^m \times I, \frac{\d x \, \d y}{\sqrt{1 - y^2}}
\right)
\ee 
and respect approach to $y = \pm 1$ end of range of $\FrQ$.

\section{Minisuperspace and slightly inhomogeneous cosmology}

The indefinite version is as before but with $t$ in place of $y$ and sign reversals within the `angular momentum' type quantities, which are now boosts.  
This is realized, for instance, in simple minisuperspace models \cite{Magic, Ryan-BI75-HH83-AMSS1}, in a conformally transformed presentation of the dynamics.    
In particular, for isotropic minisuperspace with a single scalar field, the configuration space is $\mathbb{M}^2$ and for diagonal Bianchi class A minisuperspace, it is $\mathbb{M}^3$.   

\mbox{ }

\ni In the first case, the isometry group is formed from a single `boost' $K$ and a `translational' 2-vector 
\be
P = (p_{\Omega}, p_{\phi}) 
\ee 
-- the conjugate momenta to the Misner scale variable $\Omega$ and to the scalar field variable $\phi$ -- alongside another minisuperspace 2-vector 
\be
X = (\Omega, \phi)
\ee
upon which this group acts.  
\ni These are all self-adjoint with respect to 
\be
\mbox{KinHilb} = L^2
\left(
\mathbb{M}^2, \d \Omega \, \d \phi
\right)                              \mbox{ } . 
\ee 
\be
\FrK = \mathbb{M}^2 \rtimes Poin(2)
\ee 
ensues, lying within the $\mathfrakV^*(\FrQ) \rtimes \FrG_{\scc\sa\sn}(\FrQ)$ form. 
[This is the indefinite counterpart of the familiar Heisenberg group -- here in 2-$d$ -- and $Poin$ denotes the corresponding Poincar\'{e} group.]
The objects acted upon here obey $c^2 - s^2 = 1$ and so are the hyperbolic analogue of the unit 2-vector. 

\mbox{ } 

\ni In the second case, the isometry group is the 3-$d$ Poincar\'{e} group.
This now consists of a rotation $J$ in anisotropy space, 2 boosts $K$ mixing anisotropy and Misner scale variable, and a `translational' minisuperspace 3-vector 
\be
P = (p_{\Omega}, p_{\beta_+}, p_{\beta_-},) \mbox{ } . 
\ee
Moreover, this group acts upon another minisuperspace 3-vector ($\beta_{\pm}$ are the standard anisotropy parameters \cite{Magic}) 
\be
X = (\Omega, \beta_+, \beta_-) \mbox{ } .  
\ee  
\ni These are all self-adjoint with respect to 
\be
\mbox{KinHilb} = L^2
\left(
\mathbb{M}^3, \d \Omega \, \d \beta_-\d\beta_+
\right)
\ee 
\be
\FrK = \mathbb{M}^3 \rtimes Poin(3) 
\ee
ensues, again lying within the $\mathfrakV^*(\FrQ) \rtimes \FrG_{\scc\sa\sn}(\FrQ)$ form; this is the indefinite counterpart of the 3-$d$ Heisenberg group.
The objects acted upon here obey $u^2 - v^2 - w^2 = 1$, 
e.g. $\mbox{cosh} \, \theta$, $\mbox{sinh} \, \theta \, \mbox{cos} \phi$, $\mbox{sinh} \, \theta \, \mbox{sin} \phi$, and so are the hyperbolic analogue of the unit 3-vector. 

\mbox{ } 

\ni The wave equations for the above two examples are well-known, and hence omitted from this discussion.  

\mbox{ }

\ni Let us next pass to the slightly inhomogeneous cosmology case.  
The form of the metric is again Minkowskian; 4-$d$ in fact, with line element (again in a conformally-transformed presentation) 
\be
\d s^2 = - \d \zeta_{\sn}^2 + \d v_{\sn}^2 \mbox{ } .
\ee
Here $v_{\sn} = s_{\sn}, d^{\se}_{\sn}, d^{\so}_{\sn}$.  
The $d_{\sn}$'s are tensor (T) modes, whereas the $s_{\sn}$ is the sum of the standard scalar (S) modes $a_{\sn}$ and $b_{\sn}$.
$\zeta_{\sn}$ is a `time analogue variable', defined in \cite{SIC-2}. 
Moreover, unlike in the standard Minkowski spacetime setting, this variable is now only defined on a finite interval ${\cal T}$. 
I.e. one is now restricted to a slab of this, of the form $\mathbb{R}^3 \times {\cal T}$.
Due to this, the kinematical quantization is not just the 4-$d$ version of the above minisuperspace examples, but rather in need of somewhat more elucidation.  

\mbox{ }

\ni Like for the minisuperspace examples, this involves Minkowski spacetime.
In the present case, kinematical quantization gives the algebra $\FrK$ of 18 elements [c.f. n = 3 case of Example 5) in the previous Sec]
\be
v_{i\sn}                                              \mbox{ } , \mbox{ } \mbox{ } 
\zeta_{\sn}                                           \mbox{ } , \mbox{ } \mbox{ } 
\sqrt{1 - \zeta_{\sn}^2}                              \mbox{ } , \mbox{ } \mbox{ } 
\pa/\pa v_{i\sn}                                      \mbox{ } , \mbox{ } \mbox{ } 
\sqrt{1 - \zeta_{\sn}^2}\pa/\pa \zeta_{\sn}           \mbox{ } , \mbox{ } \mbox{ }  
v_{j\sn} \pa/\pa v_{i\sn} - v_{i\sn} \pa/\pa v_{i\sn} \mbox{ } , \mbox{ } \mbox{ } 
\sqrt{1 - \zeta_{\sn}^2}\pa/\pa v_{i\sn}              \mbox{ } , \mbox{ } \mbox{ }  
\zeta_{\sn} \pa/\pa v_{i\sn}                          \mbox{ } . \mbox{ } \mbox{ } 
\ee
These are self-adjoint on 
\be
\mbox{KinHilb} = L^2
\left(
\mathbb{R}^3 \times {\cal T}, \d s_{\sn}\d d_{\sn}^o\d d^e_{\sn} \frac{\d \zeta_{\sn}}{\sqrt{1 - \zeta_{\sn}^2}} 
\right)                                                                                                             \mbox{ } .
\ee  
\ni The corresponding wave equation now involves a somewhat more complicated form of the Laplacian due to its being built out of a less straightforward explicit form for $p_t$: 
\beq
-\frac{\hbar^2}{2} 
\left\{ 
-\sqrt{1 - \zeta_{\sn}^2}  \frac{\pa}{\pa \zeta_{\sn}}  \sqrt{1 - \zeta_{\sn}^2}  \frac{\pa}{\pa \zeta_{\sn}} + \mD_{\sv_{\tn}}
\right\}
\Psi + V(\zeta_{\sn}, v_{i\sn})\Psi = E\Psi                                                                              \mbox{ } ;   
\label{SIC-KGE}
\eeq
see \cite{SIC-2} for the explicit form of $V(\zeta_{\sn}, v_{i\sn})$.  
Furthermore, (\ref{SIC-KGE}) amounts to a further correction on \cite{SIC-2}'s sketch of a wave equation, 
which in turn is a Principles of Dynamics based correction of the vacuum counterpart of \cite{HallHaw}.  

\mbox{ }

\ni N.B. that (\ref{SIC-KGE}) continues slightly inhomogeneous cosmology schemes' well-known feature of splitting into scalar, vector and tensor parts; 
each of these is coupled to the scale variable but not directly to the other.  
There is further common mathematical ground to the above suite of slightly inhomogeneous cosmology schemes.
Namely, all three are based on the same level of approximation which truncates terms at quadratic order, by which they all exhibit the mathematics of multiple harmonic oscillators.  
As such, {\it leading-order} results as regards form of quantum solutions and structure formation are expected to be unaffected by progressing along the suite, 
though differences between the quantizations are indeed expected to show up in more detailed results.

\section{Conclusion} 

\ni We considered kinematical quantization for `Cartesian' pieces of flat spaces.
These are useful as a generalization of the quite well-known $\mathbb{R}$ to $\mathbb{R}^+$ distinction, which becomes the ordinary versus affine distinction in geometrodynamics. 
These are also realized in a few cases of scaled relational mechanics (see below).
Indefinite versions are also realized in diagonal minisuperspace (well known) and modewise slightly inhomogeneous cosmology (a new problem, 
for which the current paper points to quantum-level correction terms due to kinematical quantization, in distinction to QM the equation presented in \cite{SIC-2}. 
Thus the current paper's work precedes further QM and semiclassical analysis of the last of these models \cite{SIC-3}.  

\mbox{ }

\ni The scope of kinematical quantization is moreover far greater; further areas of investigation include the following, 
and the current paper may in this regard be seen as the first of several which survey the area.

\mbox{ }

\ni 1)  Models with curved configuration spaces.
See \cite{I84} for the case of $\mathbb{S}^n$; many of the below problems also involve curved configuration spaces.  

\mbox{ } 

\ni 2) Kinematical quantization is a natural follow-up of shape geometry: 
reduced or relational configuration space quantization of RPM quantization \cite{FileR, QuadI, AMech} in \cite{AKin2}. 
This starts on quantizing the distinct 2- and 3-$d$ 3-body whole universe problems whose configuration spaces are laid out in \cite{AConfig}.  
As concrete problems, see \cite{AF, FileR} for interpretation of $\mathbb{S}^2$ as a whole-universe model consisting of 4 particles on a line, 
and \cite{+Tri, FileR} for interpretation of $\mathbb{S}^2$ as a whole-universe model consisting of a triangle of particles in 2-$d$.  
See also \cite{QuadII} for $\mathbb{CP}^2$ interpreted as a whole-universe quadrilateral of particles.
Shape-and-scale relational models, including on $\mathbb{R}^3$ and on $\mathbb{R}^3_+$, can also be found in \cite{FileR, QuadII}.  
One can expect geometrical study of \cite{AMech}'s wider range of relational mechanics' configuration spaces to lead to further range of applications of kinematical quantization.

\mbox{ }

\ni 3) Geometrical quantization can be extended from manifolds to at least some fairly well-behaved class of stratified manifolds \cite{Pflaum2, Pflaum, Brylinski, Strati}.
These require sheaf methods (see \cite{Wells} for an introduction, \cite{Nash, ID} for physical applications and \cite{Sheaves} for more in-depth accounts) 
rather than just the more familiar fibre bundle ones  
Simple concrete models of stratified manifolds start with the real-half-line with edge; 
the hemisphere with edge corresponding to the relational triangle in 3-$d$ is then the first relationally nontrivial example of a such \cite{AConfig}.
One issue with the quantization in these models understanding how each stratum can contribute its own reps, and how to interpret the totality of these reps in the quantum theory.

\mbox{ } 

\ni 4) See e.g. \cite{KR86-etc, Pflaum2, Pflaum} for gauge theory stratification, some of which reach into the interplay of this with quantization.

\mbox{ } 

\ni 5) See also \cite{Fischer70, FM96} for the presence of stratification in GR configuration spaces, \cite{I84} for a beginning on kinematical quantization consideration 
for GR in its usual geometrodynamical form and \cite{A91, AL93-Thiemann} for GR in the alternative loop form.  

\mbox{ } 

\ni 6) Finally, as regards the wider scope of models for which Isham has used Mackey's trick, 
these covered quantizing topological spaces themselves (with metric spaces considered also as a simpler case) \cite{IConcat} 
and quantization on arbitrary small categories \cite{I-Cat}.
\cite{ASoS} provides a classical precursor of the former.  

\mbox{ }

\ni{\bf Acknowledgements}. 

\mbox{ } 

\ni I thank Chris Isham for discussions over the years.  
Jeremy Butterfield for hosting in 2015-1016 and John Barrow in 2014-2015.



\begin{thebibliography}{99}

\footnotesize

\bibitem{AConfig}            E. Anderson, arXiv:1503.01507. 

\bibitem{ABook}              E. Anderson, {\it Problem of Time between Quantum Mechanics and General Relativity}, forthcoming book.

\bibitem{Arnold}            V.I. Arnol'd, {\it Mathematical Methods of Classical Mechanics} (Springer, New York 1978).  

\bibitem{Dirac}              P.A.M. Dirac, {\it Lectures on Quantum Mechanics} (Yeshiva University, New York 1964). 

\bibitem{HT92}               M. Henneaux and C. Teitelboim, {\it Quantization of Gauge Systems} (Princeton University Press, Princeton 1992).   

\bibitem{Sni}                J. \'{S}niatycki, Ann. Inst. H. Poincar\'{e} {\bf 20} 365 (1974).

\bibitem{Souriau}            J.M. Souriau, {\it Structure of Dynamical Systems. A Symplectic View of Physics} (Birkh\"{a}user, Basel 1994).  

\bibitem{BojoBook}           M. Bojowald, {\it Canonical Gravity and Applications: Cosmology, Black Holes, and Quantum Gravity} (Cambridge University Press, Cambridge 2011).   
 
\bibitem{PoT1}               K.V. Kucha\v{r}, in {\it Proceedings of the 4th Canadian Conference on General Relativity and Relativistic Astrophysics} 
                             ed. G. Kunstatter, D. Vincent and J. Williams (World Scientific, Singapore, 1992), 
%
                             reprinted as Int. J. Mod. Phys. Proc. Suppl. {\bf D20} 3 (2011); 
%
							  C.J. Isham, in {\it Integrable Systems, Quantum Groups and Quantum Field Theories} 
                             ed. L.A. Ibort and M.A. Rodr\'{\i}guez (Kluwer, Dordrecht 1993), gr-qc/9210011.

\bibitem{PoB}                K.V. Kucha\v{r}, in {\it General Relativity and Gravitation 1992},  
							 ed. R.J. Gleiser, C.N. Kozamah and O.M. Moreschi M (Institute of Physics Publishing, Bristol 1993), gr-qc/9304012; 
%
                             J.M. Pons, D.C. Salisbury and K.A. Sundermeyer, J. Phys. Conf. Ser. {\bf 222} 012018 (2010), arXiv:1001.2726;  
%
                             E. Anderson, SIGMA {\bf 10} 092 (2014), arXiv:1312.6073;  
%
                             L. Lusanna, Int. J. of Geom. Methods Mod. Phys. {\bf 12} 1530001 (2015);
%
                             B. Dittrich, P.A. Hoehn, T.A. Koslowski and M.I. Nelson, arXiv:1508.01947; 
%
							 E. Anderson, arXiv:1505.03551;  
%
                             arXiv:1604.05415.
							 

\bibitem{I84}                 C.J. Isham, in {\it Relativity, Groups and Topology {II}} ed. B. DeWitt and R. Stora (North-Holland, Amsterdam 1984).  

\bibitem{Gotay00}             M.J. Gotay, in {\it Mechanics: From Theory to Computation (Essays in Honor of Juan-Carlos Sim\'{o}} 
                              ed. J. Marsden and S. Wiggins, pp. 171-216 (Springer, New York 2000), math-ph/9809011. 

\bibitem{ABrackets}          E. Anderson, ``Brackets Structures in Theoretical Physics",                             forthcoming. 

\bibitem{Weyl27-GVH1}         H. Weyl, {\it Quantenmechanik und Gruppentheorie} (Quantum Mechanics and Group Theory) (Hirzel Verlag, Leipzig, 1928); 
%
                              H. Groenewold, Physica {\bf 12} 405 (1946). 
						  
\bibitem{Kostant-WoodhouseBook} B. Kostant, Actes, Congres Intern. Math. {\bf 2} 395 (1970); 
%
                                    N.M.J. Woodhouse, {\it Geometric Quantization} (Springer, Berlin 1980; Clarendon Press, Oxford 1991). 

\bibitem{A91}                 A. Ashtekar, {\it Lectures on Nonperturbative Canonical Gravity} (World Scientific, Singapore 1991).

\bibitem{Landsman}            N.P. Landsman, Rev. Math. Phys. {\bf 05} 775 (1993); 
%
                              {\it Mathematical Topics between Classical and Quantum Mechanics} (Springer--Verlag, New York 1998); 
%
		                      Contemp. Math. 315 (Amer. Math. Soc., Providence, Rhode Island 2002);
%
                             in {\it Handbook of the Philosophy of Physics} (Elsevier, 2005) quant-ph/0506082;
%
							 M. Kontsevich, Lett. Math. Phys. {\bf 66} 157 (2003), q-alg/9709040.  

\bibitem{RS}                  M. Reed and B. Simon {\it Methods of Modern Mathematical Physics. II. Fourier Analysis, Self-Adjointness} (Academic Press, New York 1975).  

\bibitem{Mackey}             G. Mackey, {\it Mathematical Foundations of Quantum Mechanics} (Benjamin, New York 1963).
							 
\bibitem{IConcat}            C.J. Isham, Class. Quan. Grav {\bf 6} 1509 (1989);  
%
                             in {\it Florence 1989, Proceedings, Knots, Topology and Quantum Field Theories} ed. L. Lusanna 
			    			 (World Scientific, Singapore 1989).   
%
                             ``An Introduction To General Topology And Quantum Topology", unpublished, Lectures given at Banff in 1989 (and available on the KEK archive). 
%
                             C.J. Isham, Y.A. Kubyshin and P. Renteln, Class. Quant. Grav. {\bf 7} 1053 (1990);
%
							 in {\it Moscow 1990, Proceedings, Quantum Gravity} ed M.A. Markov, V.A. Berezin and V.P. Frolov (World Scientific, Singapore 1991);  
%
                             C.J. Isham, in {\it Conceptual Problems of Quantum Gravity} ed. A. Ashtekar and J. Stachel (Birkh\"{a}user, Boston, 1991).  

\bibitem{I-Cat}              C.J. Isham,  Adv. Theor. Math. Phys. {\bf 7} 331 (2003), gr-qc/0303060; 
%
                             {\bf 7} 807, gr-qc/0304077; 
%
                             {\bf 8} 797, gr-qc/0306064. 
							 

\bibitem{IshamBook2}         C.J. Isham, {\it Modern Differential Geometry for Physicists} (World Scientific, Singapore 1999).

\bibitem{Cohom}              A. Hatcher, {\it Algebraic Topology} (Cambridge University Press, Cambridge 2001);  
%
                             J.M. Lee, {\it Introduction to Smooth Manifolds} (Springer, New York 2003);
%
                             J.G. Hocking and G.S. Young, {\it Topology} (Dover, New  York 1988); 
%
                             S. Eilenberg and N. Steenrod, {\it Foundations of Algebraic Topology} (Princeton University Press, Princeton 1952);  
%
                             R. Bott and L.W. Tu, {\it Differential Forms in Algebraic Topology} (Springer, New York 1982).  

\bibitem{Nash}               C. Nash, {\it Differential Topology and Quantum Field Theory} (Academic Press, London 1991). 

	
\bibitem{HallHaw}             J.J. Halliwell and S.W. Hawking, Phys. Rev. {\bf D31}, 1777 (1985); 
%
                              S. Wada, preprint UT Komaba85-8 (1985); 
%
                              S. Wada, Nucl. Phys. {\bf B276} 729 (1986); Erratum-ibid. {\bf B284} 747 (1987); 
%
                              Phys. Rev. {\bf D34} 2272 (1986); 
%
							  I. Shirai and S. Wada,  Nucl. Phys. {\bf B303} 728 (1988).  

\bibitem{SIC-1}               E. Anderson, arXiv:1403.7583. 

\bibitem{SIC-2}               E. Anderson, Gen. Rel. Grav. {\bf 47} 101 (2015), arXiv:1501.02443.	

\bibitem{PoT2}               E. Anderson, in {\it Classical and Quantum Gravity: Theory, Analysis and Applications}  
                             ed. V.R. Frignanni (Nova, New York 2012), arXiv:1009.2157.  
%
                             arXiv:1111.1472;
%
                             Annalen der Physik, {\bf 524} 757 (2012), arXiv:1206.2403;    
%
                             arXiv:1409.4117.               

										  
\bibitem{BB82B03FORD}        J.B. Barbour and B. Bertotti, Proc. Roy. Soc. Lond. {\bf A382} 295 (1982); 
%
                             J.B. Barbour, Class. Quant. Grav. {\bf 20} 1543 (2003), gr-qc/0211021; 
%
                             E. Anderson, Class. Quant. Grav. {\bf 25} 025003 (2008), arXiv:0706.3934.


\bibitem{FileR}              E. Anderson, arXiv:1111.1472.  

\bibitem{AMech}              E. Anderson, arXiv:1505.00488.


\bibitem{Magic}               C.W. Misner, in {\it Magic Without Magic: John Archibald Wheeler} ed. J. Klauder (Freeman, San Francisco 1972).

\bibitem{Ryan-BI75-HH83-AMSS1} M.P. Ryan, {\sl Hamiltonian Cosmology} Lec. Notes Phys. {\bf 13} (Springer, Berlin 1972); 
%
                               W.F. Blyth and C.J. Isham, Phys. Rev. {\bf D11} 768 (1975); 
%
                               J.B. Hartle and S.W. Hawking, Phys. Rev. {\bf D28} 2960 (1983);
%
                               E. Anderson, Gen. Rel. Grav {\bf 46} 1708 (2014), arXiv:1307.1916.   
										  										  

\bibitem{Affine}              C.J. Isham and A. Kakas, Class. Quant. Grav. {\bf 1} 621 (1984); 
%
                              Class. Quant. Grav. {\bf 1} 633 (1984); 
%
                              J.R. Klauder, Int. J. Geom. Meth. Mod. Phys. {\bf 3} 81 (2006), gr-qc/0507113.  
					 

\bibitem{DeWitt57}            B.S. DeWitt, Rev. Mod. Phys. {\bf 29} 377 (1957).  


\bibitem{PPSCT}               E. Anderson, Class. Quant. Grav. {\bf 27} 045002 (2010), arXiv:0905.3357.  


\bibitem{SIC-3}              E. Anderson, ``Reduced Form of Slightly Inhomogeneous Semiclassical Quantum Cosmology", forthcoming.   
					
					

\bibitem{Kendall}             D.G. Kendall, D. Barden, T.K. Carne and H. Le, {\it Shape and Shape Theory} (Wiley, Chichester 1999).  

\bibitem{AF}                  E. Anderson and A. Franzen, ``Quantum Cosmological Metroland Model", Class. Quant. Grav. {\bf 27} 045009 (2010), arXiv:0909.2436. 

\bibitem{+Tri}                E. Anderson, Gen. Rel. Grav. {\bf 43} 1529 (2011), arXiv:0909.2439.  

\bibitem{QuadI}               E. Anderson, Int. J. Mod. Phys. {\bf D23} 1450014 (2014), arXiv:1202.4186.

\bibitem{QuadII}              E. Anderson and S.A.R. Kneller, ibid. 1450052, arXiv:1303.5645.
							 
\bibitem{AKin2}               E. Anderson, ``Reduced Quantization of some 3-Body Problems", forthcoming.							 


\bibitem{KR86-etc}                W. Kondracki and J. Rogulski, Diss. Math. {\bf 250} (1986); 
%
                                  G. Rudolph, M. Schmidt and I.P. Volobuev, J. Phys. A. Math. Gen. {\bf 35} R1 (2002); 
%
                                  M. Schmidt,  Rep. Math. Phys. {\bf 51} 325 (2003).  


\bibitem{Fischer70}           A.E. Fischer, in {\it Relativity} (Proceedings of the Relativity Conference in the Midwest, held at Cincinnati, 
                              Ohio June 2-6, 1969), ed. M. Carmeli, S.I. Fickler and L. Witten (Plenum, New York 1970). 

\bibitem{FM96}                A.E. Fischer and V. Moncrief, Gen. Rel. Grav. {\bf 28}, 207 (1996).
						

\bibitem{AL93-Thiemann}      A. Ashtekar, and J. Lewandowski, J., ``Representation Theory of Analytic Holonomy $C^*$ Algebras", 
                             in {\it Knots and Quantum Gravity, Proceedings of Workshop held at UC Riverside on May 14-16, 1993, 
							 Oxford Lecture Series in Mathematics and its Applications} {\bf 1} ed. J.C. Baez (Clarendon, Oxford and OUP, New York, 1994), arXiv:gr-qc/9311010; 
%
                             T. Thiemann, {\it Modern Canonical Quantum General Relativity} (Cambridge University Press, Cambridge 2007). 
							 						


\bibitem{Pflaum}              M.J. Pflaum, {\it Analytic and Geometric Study of Stratified Spaces}, Lecture Notes in Mathematics {\bf 1768} (Springer, Berlin 2001).  

\bibitem{Pflaum2}             M.J. Pflaum, Progress in Mathematics {\bf 198} 231 (2001).

\bibitem{Brylinski}           J-L. Brylinski, {\it Loop Spaces, Characteristic Classes and Geometric Quantization} (Springer--Verlag, New York 2007). 

\bibitem{Strati}              M. Banagl, {\it Topological Invariants of Stratified Spaces} (Springer--Verlag, Berlin 2007); 
%
                              M. Kreck, {\it Differential Algebraic Topology: From Stratifolds to Exotic Spheres} (American Mathematical Society, Providence 2010); 							 
%
                              J. \'{S}niatycki, {\it Differential Geometry of Singular Spaces and Reduction of Symmetry} (Cambridge University Press, Cambridge 2013). 


\bibitem{Wells}               R.O. Wells, ``Differential Analysis on Complex Manifolds" (Springer, New York 2008).  


\bibitem{ID}                  A. Doering and C.J. Isham, in {\it New Structures for Physics}  
                              ed R. Coecke, Springer Lecture Notes in Physics 813 (Springer, Heidelberg 2011) arXiv:0803.0417. 

\bibitem{Sheaves}             B. Iversen, {\it Cohomology of Sheaves} (Springer--Verlag, Berlin 1986); 
%
                              G.E. Bredon, {\it Sheaf Theory} (McGraw--Hill, New York 1997).   

							 						 
\bibitem{ASoS}               E. Anderson, arXiv.1412.0239.
							  							  
\end{thebibliography}
\end{document}